\documentclass[twocolumn,prb,groupedaddress,showpacs]{revtex4}
\usepackage{epsfig}
\begin{document}
\renewcommand{\baselinestretch}{1}
\title{Significance of an external magnetic field on two-phonon
       processes in gated lateral semiconductor quantum dots}
\author{V.N.~Stavrou} 
\email{vstavrou@newton.physics.uiowa.edu, vstavrou@snd.edu.gr} 
\thanks{Corresponding author} 
\affiliation{Department of Physics and Astronomy, University of Iowa,
Iowa City, IA 52242, USA}
\affiliation{Division of Physics, Hellenic Naval Academy,
Hadjikyriakio, Pireus, T.K. 185 39, Greece}
\author{G.P.~Veropoulos}
\affiliation{Division of Physics, Hellenic Naval Academy, Hadjikyriakio, Pireus, T.K. 185 39
 Greece}
\begin{abstract}
Theoretical and numerical calculations of two-phonon processes on gated
lateral semiconductor quantum dots (QDs) are outlined.
A heterostructure made with two laterally coupled QDs,
in the presence of an external magnetic field, has been employed in order to study
the electron scattering rate due to two-phonon processes.
The formalism is based on the acoustic phonon modes via the unscreened deformation
potential and the piezoelectric interaction whenever the crystal lattice
lacks a center of inversion symmetry. The rates are calculated by 
using second order perturbation theory. The strong dependence of the scattering
rate on the external magnetic field, lattice temperature and QDs separation distance is 
presented.  
\newline
\newline
\it{Keywords: A. Quantum Dots; A. Semiconductors; D. Decoherence.}
\newline
\newline
* e-mail address:  vstavrou@newton.physics.uiowa.edu and tel: 0030 210 5531729
\end{abstract}
\pacs{03.67.Lx, 73.20.Dx, 85.30.Vw}
\maketitle
\date{\today}
\section{INTRODUCTION}
The gated lateral semiconductor quantum dots (QDs) in a quantum well (QW),
in which the growth direction ($z$ direction, or vertical
direction) confinement is due to the higher bandgap of the barrier material
\cite{LD,HD, Stano_2006, Baruffa_2010, Hawrylak, Hu, Khordad_2011, Sun_2012},
have been proposed for use as quantum bits (qubits) in 
quantum computer architecture several times.
Decoherence due to single electron confinement within a coupled QDs structure,
which plays the role of qubits (QBs),
has two important channels. The first channel is the Coulomb interaction to the background
charge fluctuation and the second is the electron-phonon interaction.
Furthermore, physical properties of semiconductor-based QBs relevant 
to single electron spin and 
charge degree of freedom have been a subject of theoretical studies.\cite{Hu}
Notice that for charge qubits there is only a single electron in a double dot,
in contrast with spin qubits, where each quantum dot has one electron and
the double dot is only for two-qubit operations.\cite{LD,HD}
\par
The effects of charge decoherence due to electron-phonon interactions
are of crucial importance in semiconductor-based quantum 
computer architecture.
In our recent work,\cite{Stavrou_Hu2005, Stavrou_Veropoulos}
we have shown that the scattering rates due to 
electron-acoustical phonon interactions and the dephasing rates due to the coupling of 
electrons to acoustical and optical phonons strongly depend on the interdot distance 
and the strength of the electron confinement. 
The multiphonon processes \cite{Two_phonons} among other scattering mechanisms
(e.g. one phonon process, electron-electron, spin-phonon interactions etc)
can describe and measure the decoherence in QDs. 
Earlier theoretical work,  in multiphonon processes in single three dimensional QD made with GaAs
in which the electron confinement potential is assumed to be isotropic and parabolic, was 
reported two decades ago.\cite{Inoshita92}
In their theoretical work, they used a few possible processes using longitudinal acoustical (LA)
and longitudinal optical phonons (LO)
which were mainly described by bulk phonon approximation.\cite{Essex}
This theory serves laser nanotechnology interests and handles the photoluminescence degradation 
in small QDs.
According to the best knowledge of the authors, the relaxation rates
(source of charge decoherence in qubits and in optoelectronic devices) 
due to multiphonon processes in laterally coupled QDs in the presence
of an external magnetic field and for a range of operating temperature
have not been reported.
\par
In this paper, our study related to two phonon
processes in coupled QDs under the existence of an external magnetic field
and their role in charge decoherence is presented. 
Starting with section II, we firstly give a theoretical motivation of the electron wavefunctions
and phonon model which describes the deformation and piezoelectric types of electron-phonon
interactions. Secondly, we present the equation of electron scattering rates due to the 
second order perturbation term which describes the two-phonon processes.
In section III, we show the relaxation rates due to two-phonon processes and we
discuss their dependence on several configurations.
Lastly, section IV, presents a summary of our results and future implementations.
%
%
\section{THEORY}
%
We consider a heterostructure composed of two laterally coupled QDs. 
In order to calculate the electron states within the coupled system,
we have used a one-band effective mass approximation. 
The Hamiltonian which describes the single-electron motion \cite{Jacak}
which is confined in laterally coupled QDs is given by 
\begin {equation}
\label{hamil}
\hat{\mathcal{H}} = \hat{\mathcal{H}}_{\parallel} + \hat{\mathcal{H}_{z}}
\end {equation}
where the lateral motion of electron is decoupled from the one
along the quantum well growth (z-axis)\cite{Stavrou_Hu2005, Bockel}.
The external magnetic field is applied along the z-axis 
$({\bf B} =B {\bf \hat e_{z}})$,
and as a result the magnetic vector potential ${\bf A}$ could be
given as
\begin {equation}
\label{vector_potential}
{\bf A}=B\left( -y{\bf \hat e_{x}} + x{\bf \hat e_{y}}\right)/2
\end {equation}
The Hamiltonian operators for the lateral directions and z-direction
have been considered as
\begin {equation}
\label{hamil_xy}
\hat{\mathcal{H}}_{\parallel} = \frac{\hat{p}^{2}}{2m^{\ast}}
                              + \frac{1}{2}m^{\ast}\omega^{2}r_{\parallel}^{2}
			      - \frac{1}{2}\omega_{c}{\bf {L}}_{z} 
\end {equation}
\begin{equation}
\label{hamil_z}
\hat{\mathcal{H}_{z}} =
-\frac{\hbar}{2}\partial_{z}\frac{1}{m^{\ast}(z)}\partial_{z}
                        +
V_{0}\Theta\left(\left|z\right|-L_{0}\right) \,.
\end{equation}
where ${\bf L_{z}}$ is the operator of the $z$ component of the angular momentum,
$m^{\ast}(z)$ is the electron effective mass, $V_{0}$ is the offset
between the band edges of the GaAs well and the AlGaAs barrier, 
$\Theta$ is the Heaviside step function, $\hat{p}$ is the quantum 
mechanical operator of momentum, $\omega_{0}$ is a parameter 
describing the strength of the confinement in x-y plane,
$\omega = Be/m^{*}$ and $\omega^{2}=\omega^{2}_{0}+(\omega_{c}/2)^{2}$.
\par
According to Eq.~(\ref{hamil}), electron wavefunction
can be given by the following envelope function,
\begin {equation}
\label{envelope}
\psi(\bf{r}) = \psi_{\parallel}\left(\bf{r}_{\parallel}\right)
               \psi_{\it{z}}\left(\it{z}\right)
\end {equation}

In our investigation, we have only considered the ground state
wavefunction along the QW growth 
and the wavefunction along the lateral direction is
given by Fock-Darwin states.
Following the same procedure as \onlinecite{Stavrou_Hu2005}, we have considered that the external confining
potential for the electron within two QDs structure is given by
\begin {equation}
\label{Vc}
V_{c} = \frac{1}{2}m^{\ast}\omega_{0}^{2}~min\{ \left(x-\alpha\right)^{2}+y^{2},~
                                                    \left(x+\alpha\right)^{2}+y^{2} 
                                                    \}
\end {equation}
where $\alpha$ is the separation distance of the dots.
The electron wavefunction of the coupled QD structure could be described by
\begin {equation}
\label{wavefunction}
\Psi(\bf{r}) = \Psi_{\|}\left(\bf{r_{\|}}\right)
                \psi_{\it{z}}\left(\it{z}\right)
\end {equation}
where the single electron wavefunction for the parallel plane is given by: 
\begin {equation}
\label{superposition}
\left |\Psi_{\|}\right> = \sum_{k}{C_{k}\left |\psi_{\|,L}^{k}\right> + 
                              D_{k}\left |\psi_{\|,R}^{k}\right>}
\end {equation}
A numerical scheme has been employed in order to calculate the total wavefunction
in the parallel plane of the coupled dot system.
In low dimensional structures, the electrons interact with acoustical and optical phonons. 
The optical phonons do not have any contribution to 
electron scattering rates due to the small electron energy
splitting. Therefore, only the acoustical phonons contribute to the
relaxation rates. In this work, we calculate the electron scattering 
rate which is caused due to deformation potential and piezoelectric
acoustic phonon interaction \cite{Mahan, MahanPAP1, Bruus1993}.    
The Hamiltonian which describes these interactions 
is given by: 
\begin {equation}
\label{phonons1}
H = \sum_{\bf q} \left( \frac{\hbar}{2 \rho_m V \omega_{\bf q}}
\right)^{1/2} {\mathcal{M}}({\bf q})
\rho({\bf q}) (a_{\bf q}+a_{-\bf q}^{\dagger}) \,,
\end {equation}
The term $\mathcal{M}({\bf q})$, which includes both the deformation
and the piezoelectric interaction for zincblende crystals, is defined by
\begin {equation}
\label{phonons2}
\mathcal{M}({\bf q}) = D\left| \bf{q} \right| + {\it i}\mathcal{M}^{pz}_{\lambda}(\hat {q})
\end {equation}
with
\begin {equation}
\label{PZ}
{\mathcal{M}}^{pz}_{\lambda} (\hat{\bf q}) = 2 e \ e_{14}\left( \hat{q}_{x}
\hat{q}_{y} \xi_{z} + \hat{q}_{y} \hat{q}_{z} \xi_{x} + \hat{q}_{x}
\hat{q}_{z} \xi_{y} \right)
\end {equation}
In Eqs.~(\ref{phonons1}-\ref{PZ}), $\rho_m$ is the mass density of the
host material, $\omega_{\bf q}$ is the frequency of the phonon mode
with wavevector ${\bf q}$,
$V$ is the volume of the sample, $a_{\bf q}$ and $a_{-\bf
q}^\dagger$ are phonon annihilation and creation operators, $\rho({\bf
q})$ is the electron density operator, $D$ denotes the
deformation potential, $e_{14}$ is the piezoelectric constant
and $\xi$ is the polarization vector. All values of the above mentioned
parameters used in our calculations have been taken
from Ref. \onlinecite{Bruus1993}.
\par
The last part of our theoretical formalism is the calculation of the electron
scattering rates due to two-phonon processes. Considering only LA phonons, 
the scattering rates (second order perturbation theory) are given
by the following equations
\begin {eqnarray}
\label{fermi1}
\Gamma_{++}  &=&  \frac{\pi}{\hbar}
\sum_{\bf{q}}
\sum_{\bf{k}}
\left| \sum_{s} 
\left(\frac{M_q^{is}~M_k^{sf}}{E_{i}-E_{s}-E_{q}}+\frac{M_k^{is}~M_q^{sf}}{E_{i}-E_{s}- E_{k}}
\right)
\right|^{2} 
\nonumber\\ & &
\left(N_{q}+1\right) 
\left(N_{k}+1\right) 
\delta\left(E_{i}-E_{f} - E_{q} - E_{k} \right)
\end {eqnarray}
\begin {eqnarray}
\label{fermi2}
\Gamma_{+-}  &=&  \frac{2\pi}{\hbar}
\sum_{\bf{q}}
\sum_{\bf{k}}
\left| \sum_{s} 
\left(\frac{M_q^{is}~M_k^{sf}}{E_{i}-E_{s}-E_{q}}+\frac{M_k^{is}~M_q^{sf}}{E_{i}-E_{s}+ E_{k}}
\right)
\right|^{2} 
\nonumber\\ & &
N_{k} \left(N_{q}+1\right) 
\delta\left(E_{f}-E_{i} - E_{q} + E_{k} \right)
\end {eqnarray}
\begin {eqnarray}
\label{fermi3}
\Gamma_{--}  &=&  \frac{\pi}{\hbar}
\sum_{\bf{q}}
\sum_{\bf{k}}
\left| \sum_{s} 
\left(\frac{M_q^{is}~M_k^{sf}}{E_{i}-E_{s}+E_{q}}+\frac{M_k^{is}~M_q^{sf}}{E_{i}-E_{s}+ E_{k}}
\right)
\right|^{2} 
\nonumber\\ & &
N_{q} N_{k} 
\delta\left(E_{f}-E_{i} + E_{q} + E_{k} \right)
\end {eqnarray}
where the indices ${++}$, ${--}$, ${+-}$ represent the emission of two phonons (LA+LA),
the absorption of two phonons(-LA-LA) and the emission of one phonon and absorption
of one phonon (LA-LA or -LA+LA) respectively. 
$M_q^{sf}$ stands for the electron-phonon matrix elements 
where the index i (f) corresponds to qubit electron first excited state
(ground state) and s stands for the intermediate electronic states.
The other elements are taken by changing the proper suffixes.
$N_{k}$ ($N_{q}$) is the Bose distribution function referring to phonons 
with energy $E_k=\hbar\omega_{k}$ ($E_q=\hbar\omega_{q}$).
Note that the summation over s excludes the initial and final states.  
The integrals which are included in Eqs.~(\ref{fermi1}-\ref{fermi3})
by transforming the summations to integrations, have been calculated
by Monte Carlo code. 
%
\section{RESULTS}
\begin{figure}[]
\begin{center}
\includegraphics[trim= 0 0 0 -45, width=3.0in]{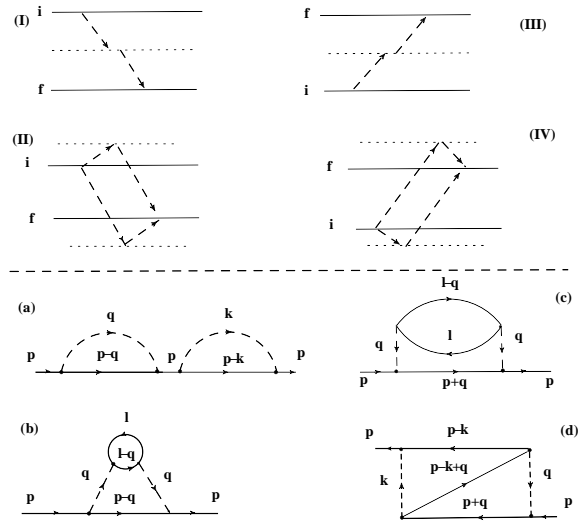}
\protect\caption{The Feynman graphs for the four different transitions related to two-phonon processes.}
\label{fig_0_two_phonons}
\end{center}
\end{figure}

%
%
Fig. \ref{fig_0_two_phonons} shows all possible scattering processes concerning the
electron transitions due to second order contributions associated to acoustic phonons.
The transitions described by eqs. (\ref{fermi1}) and (\ref{fermi3}) are presented in 
Fig. \ref{fig_0_two_phonons}-I (Fig. \ref{fig_0_two_phonons}-b) and 
Fig. \ref{fig_0_two_phonons}-III (Fig. \ref{fig_0_two_phonons}-c) respectively.
It is worth mentioning
that eq. (\ref{fermi2}) creates two different transitions as illustrated in 
Fig. \ref{fig_0_two_phonons}-II (Fig. \ref{fig_0_two_phonons}-b) and 
Fig. \ref{fig_0_two_phonons}-IV (Fig. \ref{fig_0_two_phonons}-d). 
Using the results of the second order perturbation theory 
( Eqs.~\ref{fermi1}-~\ref{fermi3}),
we estimate the relaxation rates for an electron which relaxes to ground state via the 
two phonon processes.
\par
\begin{figure}[]
\begin{center}
\includegraphics[trim= 0 0 0 -45, width=3.0in]{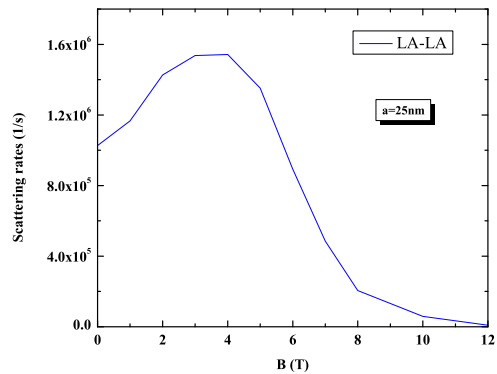}
\protect\caption{Relaxation rates of an electron due to
multiphonon process LA-LA versus the external magnetic field
$B$. The confinement strength is $\hbar\omega= 3~meV$, lattice temperature
$T=1~K$ and QW width $2L_{z}=10~nm$.}
\label{fig_2}
\end{center}
\end{figure}

In Fig.  \ref{fig_2}, we present the relaxation rates for
the case of the emission of a LA phonon and the absorption of a LA phonon (LA-LA),
as a function of an external magnetic field.
Increasing the magnetic field in the range of 0-12 T, the electron wavefunctions get 
the largest value (resonance value) at B~3.7 T and as a result the matrix elements 
involved in the two-phonon scattering process increase. Furthermore, the increasing 
number of phonon modes (for B = 0-4 T) that can be involved in the relaxation process, 
increases the scattering rates.
For larger values of the magnetic field (B = 4-12 T) the rates decrease due to the 
fact that the electron wavefunctions move away from the resonance value and due to 
the decreasing number of phonon density of phonons \cite{Stavrou_Hu2005}. 
The dependence of matrix elements on wavevector is related to
$\sqrt{q}$ for the deformation potential, while for the piezoelectric 
coupling it is related to $1/\sqrt{q}$. 
It is worth mentioning that although we have included both
deformation and piezoelectric interactions in our phonon
description, it is impossible to separate their contribution
to the electron relaxation rates though two-phonon processes
because of the dependence of matrix elements on the
interaction strength (eq.\ref{phonons2}).  
\par
\begin{figure}[]
\begin{center}
\includegraphics[trim= 0 0 0 -45, width=3.0in]{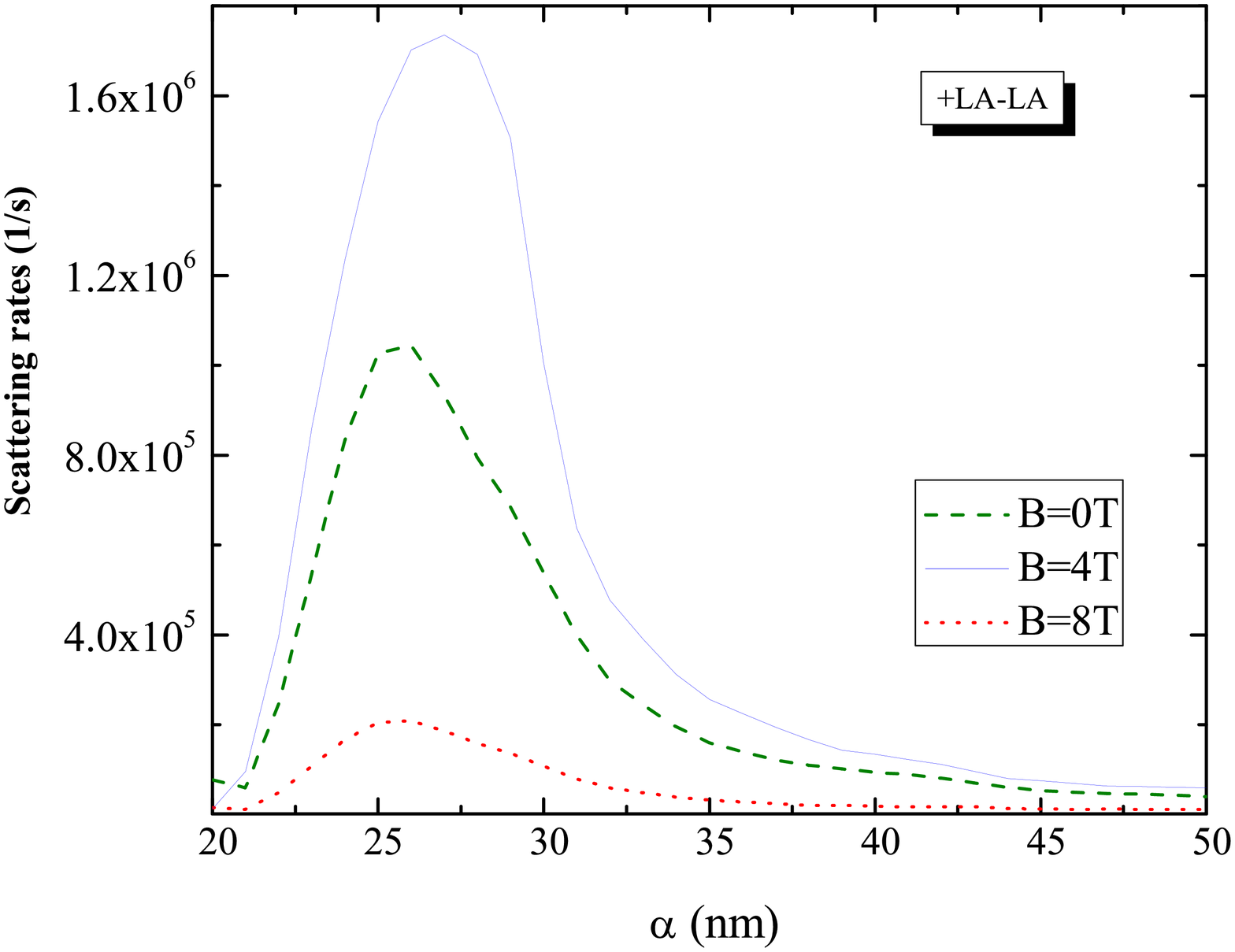}
\protect\caption{Relaxation rates of an electron due to
multiphonon process LA-LA versus the half interdot distance $\alpha$.
The lattice temperature is fixed to $T=1~K$ and QW width is fixed to $2L_{z}=10~nm$.}
\label{fig_3}
\end{center}
\end{figure}

The dependence of the electron scattering rates on the half of the
interdot separation distance for a range of magnetic field values is presented in Fig. 
\ref{fig_3}. It is obvious that for $B=4~T$ the rates 
have the largest value because the electron-phonon matrix elements get the
highest value at magnetic field $3.7~T$.
Increasing the interdot distance, the rates emerge peaks due to the increasing 
number of phonon density of phonons and due to the matrix elements enhancement.
For large interdot distance, the scattering rates decrease because of the
decreasing number of phonon modes that can be involved in the scattering rates.

%
\begin{figure}[]
\begin{center}
\includegraphics[trim= 0 0 0 -45, width=3.0in]{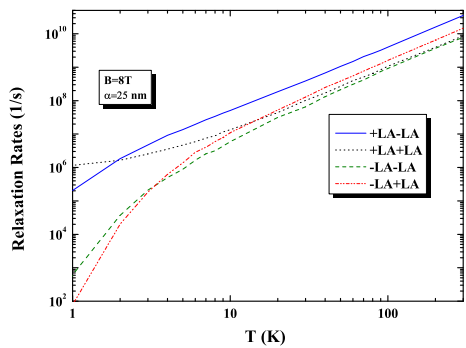}
\protect\caption{Relaxation rates of an electron due to four different 
 multiphonon processes versus the lattice temperature.
The magnetic field is fixed to $B=8T$ confinement strength is $\hbar\omega= 3~meV$,
half interdot distance $\alpha = 25~meV$ and QW width $2L_{z}=10~nm$.}
\label{fig_4}
\end{center}
\end{figure} 
\par
Fig. \ref{fig_4} shows the calculated relaxation rates as a function of the lattice
temperature for all possible two phonon processes
for a non zero applied external magnetic field. 
The first feature is that there are crossovers between the rates of different 
phonon processes due to different phonon population factors related to
Bose distribution function (see Eqs.~(\ref{fermi1}-\ref{fermi3}) ).
Secondly, the rates increase rapidly by increasing the lattice temperature from 
low temperatures-which are generally the operating temperature of the 
devices made with low dimensional structures-up to $300~K$.
The rise of the rates reaches almost 8 order of magnitude difference
(-LA+LA) for the examined edges of temperature interval. This type of behavior is
because of differing phonon population factors as the lattice temperature
increases. 
Lastly, at room temperature the processes +LA+LA and -LA-LA
have almost the same scattering rates due to the same phonon distribution.
The same feature also exists for the rates corresponding to the +LA-LA and -LA+LA
processes for very large temperature.
%
%
%
%
\section{CONCLUSIONS}
%
We have researched the decoherence channel due to electron-phonon interaction by
studying the two-phonon processes on gated lateral semiconductor QDs. 
We have studied the electron coupling 
to acoustical phonons through a deformation potential and piezoelectric interaction
and found a strong dependence of relaxation rates on the external magnetic field,
the separation distance and the lattice temperature.
Although the electron
relaxation rates can have very tiny values for the $mK$ range, 
the increase of temperature can increase the rates to large values.
Lastly, according to the best of the authors knowledge, such experiments
related to charge decoherence for our geometric configuration in the presence
of an external magnetic field and large operating temperature have not been
reported in order to compare them with our theoretical results.
%
\section{ACKNOWLEDGMENT}
%
The work is supported in part by NSA and ARDA under ARO contract
No.~DAAD19-03-1-0128. The author V.N.S. would like to
thank Prof. X. Hu for useful discussions.
%

%
%
%
\end{document}